\documentclass[%
reprint,
superscriptaddress,
nofootinbib,
 amsmath,amssymb,
 aps,
]{revtex4-2}

\usepackage[dvipsnames]{xcolor}
\usepackage{svg}
\usepackage{graphicx} 
\usepackage{dcolumn} 
\usepackage{bm} 
\usepackage{xurl} 
\begin{document}

\preprint{APS/123-QED}

\title{Strong Field Ionization of Water:\\
Nuclear Dynamics Revealed by Varying the Pulse Duration}
\author{A. J. Howard}
\affiliation{Department of Applied Physics, Stanford University, Stanford, CA 94305}
\affiliation{Stanford PULSE Institute, SLAC National Accelerator Laboratory\\
2575 Sand Hill Road, Menlo Park, CA 94025}

\author{C. Cheng}
\affiliation{Department of Physics, Stony Brook University, Stony Brook, NY 11794}

\author{R. Forbes}
\affiliation{Stanford PULSE Institute, SLAC National Accelerator Laboratory\\
2575 Sand Hill Road, Menlo Park, CA 94025}
\affiliation{Department of Physics, Stanford University, Stanford, CA 94305}

\author{G. A. McCracken}
\affiliation{Department of Applied Physics, Stanford University, Stanford, CA 94305}
\affiliation{Stanford PULSE Institute, SLAC National Accelerator Laboratory\\
2575 Sand Hill Road, Menlo Park, CA 94025}

\author{W. H. Mills}
\affiliation{Department of Chemistry and Physics, University of Mary Washington, Fredericksburg VA 22401}

\author{V. Makhija}
\affiliation{Department of Chemistry and Physics, University of Mary Washington, Fredericksburg VA 22401}

\author{M. Spanner}
\affiliation{National Research Council of Canada, 100 Sussex Drive, Ottawa K1A 0R6, Canada}
\affiliation{Department of Physics, University of Ottawa, Ottawa K1N 6N5, Canada}

\author{T. Weinacht}
\affiliation{Department of Physics, Stony Brook University, Stony Brook, NY 11794}

\author{P. H. Bucksbaum}
\altaffiliation{Corresponding author. Email: phbuck@\hspace{0.25mm}stanford.edu}
\affiliation{Department of Applied Physics, Stanford University, Stanford, CA 94305}
\affiliation{Stanford PULSE Institute, SLAC National Accelerator Laboratory\\
2575 Sand Hill Road, Menlo Park, CA 94025}
\affiliation{Department of Physics, Stanford University, Stanford, CA 94305}

\begin{abstract}
\noindent Polyatomic molecules in strong laser fields can undergo substantial nuclear motion within tens of femtoseconds. Ion imaging methods based on dissociation or Coulomb explosion therefore have difficulty faithfully recording the geometry dependence of the field ionization that initiates the dissociation process. Here we compare the strong-field double ionization and subsequent dissociation of water (both H$_2$O and D$_2$O) in 10-fs and 40-fs 800-nm laser pulses. We find that 10-fs pulses turn off before substantial internuclear motion occurs, whereas rapid internuclear motion can take place during the double ionization process for 40-fs pulses. The short-pulse measurements are consistent with a simple tunnel ionization picture, whose predictions help interpret the motion observed in the long-pulse measurements.
\end{abstract}

\maketitle
\section{\label{sec:Introduction}Introduction}

    The momentum distribution of ionic fragments following rapid stripping of valence electrons is often used to reconstruct the nuclear geometry of molecules immediately before dissociation \cite{vager_coulomb_1989}.
    This technique was originally developed with a fast beam of molecules passing through a thin foil to strip away the electrons, but subsequent measurements use intense ultrafast lasers to remove the electrons \cite{stapelfeldt_time-resolved_1998}. 
    The critical difference between these two methods is that the stripping fields of a foil are random, and therefore do not depend on the orientation of the molecules, whereas the laser field is polarized and thus laser-driven ionization can depend strongly on the orientation of the molecule with respect to the polarization.  

    In the tunneling theory of Strong Field Ionization (SFI), molecular orbitals have preferred orientations that maximize ionization probability with respect to a linearly polarized laser field. Ionization or excitation that prefers a particular alignment based on orbital shape is called geometric alignment. Through this mechanism, an isotropically oriented ensemble of molecules can exhibit strongly anisotropic momentum distributions \cite{stapelfeldt_colloquium_2003,posthumus_dynamic_1998,alnaser_effects_2005,pavicic_direct_2007}.

   The strong fields required for laser-driven multiple ionization, on the order of $\mathrm{V/\text{\AA}}$, can also distort the results by inducing ultrafast motion during ionization 
   \cite{ellert_observing_1998,legare_laser_2005,legare_laser_2006,ibrahim_H2_2018}. One manifestation of this is dynamic alignment, in which an anisotropic polarizability leads to a torque in the presence of a strong field \cite{stapelfeldt_colloquium_2003}. 
   In SFI experiments on isotropically oriented small molecules with small moments of inertia, dynamic alignment by a linearly polarized laser field drives molecules into rotation and/or bond-angle distortion as the molecules reach a favorable alignment with respect to the polarization \cite{posthumus_double-pulse_1998,posthumus_dynamic_1998,miyazaki_femtosecond-laser-induced_2004,madsen_application_2014}.

    For lighter molecules, both geometric and dynamic alignment play a significant role in femtosecond strong-field ionization, and must both be considered to understand the resulting angle-dependent ionization yields 
    \cite{posthumus_double-pulse_1998,posthumus_dynamic_1998}. 
    An important parameter that helps to disentangle these two effects is the ionizing pulse duration \cite{legare_laser_2005,banerjee_dynamic_1999,koh_ionization_2020,rosca-pruna_spatial_2001}. 
    When the pulse duration is much shorter than the time-scale of dynamic alignment in the molecule, geometric alignment dominates the angle dependence of the ionization yield. Conversely, when the pulse duration is much longer than this, dynamic alignment can obscure the effects of geometric alignment \cite{mccracken_ionization_2020}.
   
\begin{figure*}
\includegraphics[width=17.2cm, trim={2.25cm 0cm 1.95cm 0cm}]{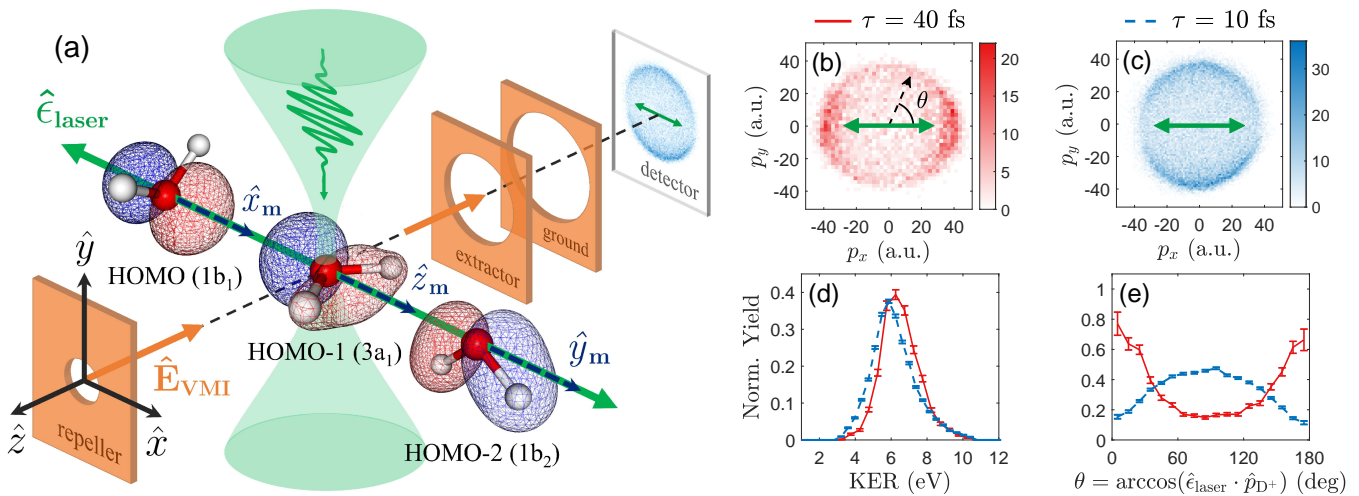}
    \caption{(a)~A schematic rendering of the ionization process and detection geometry. A focused femtosecond laser pulse (green), with polarization $\hat{\epsilon}_{\mathrm{laser}}$ along the $\hat{x}$-direction, removes electrons from the three highest occupied molecular orbitals (HOMO, HOMO-1, and HOMO-2) of neutral water. Preferential ionization of each orbital in a particular orientation with respect to the polarization induces geometric alignment as indicated by the molecular axis ($\hat{x}_{\mathrm{m}}$/$\hat{y}_{\mathrm{m}}$/$\hat{z}_{\mathrm{m}}$) that is parallel to $\hat{\epsilon}_{\mathrm{laser}}$. After ionization, ionic fragments are accelerated toward the detector screen by an electrostatic lens stack (orange) that applies a constant electric field $\hat{E}_{\mathrm{VMI}}$ in the $\hat{z}$ direction. The net result of this scheme is an image that forms on the detector plane capturing the $\hat{x}$- and $\hat{y}$-momenta of each fragment as well as a distribution of fragment arrival times indicative of the $\hat{z}$-momentum. (b)~The $\hat{x}$ and $\hat{y}$ momentum distribution of D$^+$ ions captured in coincidence with OD$^+$, following ionization with a 40-fs pulse. The laser polarization axis is drawn as a green arrow. (c)~The same distribution as in (b), but following ionization with a 10-fs pulse. (d)~The kinetic energy release (KER) distribution plotted for both pulse durations. (e)~The angular distribution of D$^+$ ions with $\theta$ defined as the angle between the D$^+$ momentum and the laser polarization axis, as drawn in panel (b).}
\label{fig:Schematic}
\end{figure*} 
   
   Previous work has highlighted that ultrafast molecular motion can align the momenta of dissociating fragment ions in multiply-ionized water \cite{mccracken_ionization_2020, mccracken_geometric_2017, sanderson_geometry_1999,legare_laser_2005, tong_post_2005, zhao_strong-field-induced_2019}. Water is a ten-electron molecule with three filled valence molecular orbitals (HOMO, HOMO-1, and HOMO-2) closely analogous to the degenerate $\textit{n}$=2, $\ell$=1 electron orbitals of neon: 2p$_x$, 2p$_y$, and 2p$_z$. The redistribution of electron density among two hydrogens and a non-collinear oxygen breaks both the degeneracy and spherical atomic symmetry so that each orbital is distinguishable (see Fig.~\ref{fig:Schematic}(a)).
 
  The SFI of water has recently been demonstrated to obey a simple tunneling picture \cite{cheng_momentum-resolved_2020}, where a strong linearly polarized field selectively ionizes electrons whose nodal plane is perpendicular to the polarization \cite{kjeldsen_role_2003,stapelfeldt_colloquium_2003,mccracken_geometric_2017}.
  This leads to simple predictions of the orientation dependence for strong-field ionization, otherwise known as geometric alignment:
  The HOMO (1b$_1$) orbital should preferentially ionize when the polarization is aligned with the $\hat{x}_{\mathrm{m}}$ molecular axis;
  HOMO-1 (3a$_1$) ionizes for polarization along $\hat{z}_{\mathrm{m}}$, and HOMO-2 (1b$_2$) along $\hat{y}_{\mathrm{m}}$. These three orbitals as well as their preferred alignments with respect to the polarization are represented schematically in Fig.~\ref{fig:Schematic}(a). Such predictions have been supported by both experimental evidence and theoretical strong-field angle-dependent ionization calculations \cite{mccracken_geometric_2017,cheng_momentum-resolved_2020}.
  
  Simple tunneling considerations for each orbital lead to straightforward predictions of the angle-dependent ionization yields for the low-lying states of both the monocation and dication. For example, the doublet ground state of the monocation, D$_0$~(1b$_1$)$^{-1}$, should have a maximum ionization yield for polarization along $\hat{x}_{\mathrm{m}}$. The triplet ground state of the dication, T$_0$~(1b$_1$)$^{-1}$(3a$_1$)$^{-1}$, should have a maximum in its ionization yield somewhere within the $\hat{x}_{\mathrm{m}}\hat{z}_{\mathrm{m}}$ plane. These angle-dependent ionization yields are what produce geometric alignment in the monocation and dication populations. 
  
  Similar considerations lead to predictions for the dependence of dissociation channels on the occupied molecular orbitals. For example, two-body decays of the dication are expected following the removal of an electron from HOMO or HOMO-1; while three-body decays are expected following the removal of a HOMO-2 electron \cite{gervais_h2o2_2009,streeter_dissociation_2018,reedy_dissociation_2018}. Significant deviations from such predictions in the tunnel-ionization regime indicate the presence of laser-induced effects such as alignment or internuclear motion in the monocation or the dication \cite{mccracken_ionization_2020}.
   
   The polarizability of water is very nearly isotropic in its neutral ground state \cite{spelsberg_static_1994}, so laser-induced alignment is not expected prior to ionization. However, ultrafast molecular dynamics in the monocation and dication could play a role in dissociation following the second ionization to form H$_2$O$^{2+}$. Laser-driven dynamic alignment can also be significant if the ionizing pulse duration significantly exceeds the $\sim$10~fs needed to realign the H-H axis to the laser polarization before dissociation \cite{mccracken_ionization_2020}.
   
   Here we compare the dissociation patterns following strong-field ionization of water with laser pulse durations of $\tau$~=~40~fs and 10~fs, and with comparable peak intensities of I$_0$~=~600~TW/cm$^2$ and 400~TW/cm$^2$, respectively. In each case, all three stable isotopes of water (H$_2$O, D$_2$O, and HOD) were doubly ionized and the dissociation of the resultant dication (H$_2$O$^{2+}$/D$_2$O$^{2+}$/HOD$^{2+}$) was observed using a velocity map imaging (VMI) apparatus, yielding the full 3D vector momenta of all fragments. For the purposes of data fidelity, the following sections will focus primarily on the experimental results recorded for D$_2$O$^{2+}$.

\section{\label{sec:Experiment}Experimental Methods}
   The momentum distributions of ions following the double ionization of water with long (40-fs) and short (10-fs) pulses were obtained using two separate VMI systems. Each has been described previously, so only a brief summary is provided here.
   
   The long-pulse experiment employs linearly polarized Ti:Sapphire laser pulses with a central wavelength of 800~nm and a pulse duration of $\tau$~=~40~fs full width at half maximum (FWHM)
   \cite{mccracken_geometric_2017,mccracken_ionization_2020}.
   The 1-kHz laser is focused to an intensity of 600~TW/cm$^2$ at the interaction region of a vacuum chamber (base pressure of $10^{-10}$~mbar), where it intersects a skimmed effusive molecular beam of the stable isotopes of water (H$_2$O, D$_2$O, and HOD). 
   The \textit{in situ} intensity was calibrated by SFI of argon, using a procedure outlined in Ref.~\cite{bryan_atomic_2006}. As shown schematically in Fig.~\ref{fig:Schematic}(a), the interaction region sits at the center of an electrostatic lens stack, allowing the ionic products of SFI to be collected in a VMI spectrometer. The apparatus uses a Roentdek Hexanode detector, with a time resolution of $<$1~ns \cite{jagutzki_multiple_2002}, 
   to capture the full 3D vector momenta of the ions produced in the interaction region.
   
   The short pulse experiment uses a 780-nm 30-fs 1-kHz Ti:Sapphire system 
   \cite{zhao_coincidence_2017,cheng_electron_2019}. Output pulses from this laser were spectrally broadened in an argon gas filament and recompressed to approximately 10~fs (FWHM) using chirped mirrors and an acousto-optic pulse shaper \cite{dugan_high-resolution_1997}. The recompressed, linearly polarized pulses are then focused into the interaction region of a vacuum chamber, held at a base pressure of $10^{-10}$~mbar, by an internally mounted $f$ = 5~cm concave spherical mirror. The stable isotopes of water are introduced into the VMI chamber via a skimmed effusive molecular beam. Here, an electrostatic lens stack can be rapidly switched from positive to negative voltages \cite{lehmann_velocity_2012} in order to measure both electrons and ions from each molecule. Directly following ionization, the lens stack accelerates free electrons toward a detector consisting of a microchannel plate (MCP) detector, phosphor screen, and TimePix3 camera 
   \cite{fisher-levine_timepixcam_2016,nomerotski_characterization_2017}. Less than 25~ns later, the voltages are reversed and the positive ions are accelerated toward the same detector. With a time resolution of $\sim$1~ns, this detector is able to resolve the 3D vector momenta of all ions, as well as the 2D vector momenta of the electrons (projected onto the plane of the detector). In this work, only the ion measurements are considered; the electron data has been considered in a previous study \cite{cheng_momentum-resolved_2020}.
   Importantly, the rapid voltage switching can introduce slight distortions to the measured momenta of the positive ions. In this experiment, the only substantial distortions were found in the momentum distributions of the lightest fragment, H$^+$. Due to this effect, we examined both D$_2$O and H$_2$O to ensure the validity of our conclusions. 
   
\section{\label{sec:Results}Results and Discussion}

\subsection{\label{subsec:Two-Body}Two-Body Fragment Momentum Distributions}   

  The relative contributions of geometric and dynamic alignment in the double ionization of water can be observed in the fragment ion momentum distribution of the D$_2$O$^{2+}$ two-body breakup channel, where D$^+$ and OD$^+$ are produced. In this channel, the detector records the full vector momentum for each fragment ion: $p_x$ and $p_y$ are determined by the position of the hit in the detector plane, and $p_z$ is determined by the time-of-flight (ToF) (see Fig.~\ref{fig:Schematic}(a)). The ToF is also used to identify the species of the ion using its unique mass-to-charge ratio~\cite{ullrich_recoil-ion_2003}. Figs.~\ref{fig:Schematic}(b) and (c) are two-dimensional projections along $p_z$ of the momentum distribution in the $p_x p_y$ plane of D$^+$ ions detected in coincidence with an OD$^+$ fragment, for long and short pulses respectively.
The angular distributions with respect to the laser polarization (drawn as a green arrow) are visibly different. 
The 40-fs pulse leads primarily to dissociation parallel to the polarization axis whereas the 10-fs pulse leads to dissociation perpendicular to the laser polarization. 
The sum of the kinetic energy of these fragments yields the dissociation kinetic energy release (KER). 
The KER distributions for D$^+$/OD$^+$ is plotted in Fig.~\ref{fig:Schematic}(d). Note that the long-pulse data is slightly higher in KER, despite the fact that realignment effects would typically reduce the energy of dissociating fragments \cite{bucksbaum_softening_1990}. This will be discussed further in a later subsection. Calculating the relative angle between the vector momentum of the D$^+$ ion and the laser polarization yields Fig.~\ref{fig:Schematic}(e). This plot reaffirms what was noted in comparing Figs.~\ref{fig:Schematic}(b) and (c): the dissociation is parallel to the polarization for long pulses and perpendicular to the polarization for short pulses.
  
The difference in the dissociation distributions for the short and long pulse experiments suggests some evolution on the potential energy surfaces of the monocation or dication before dissociation. If this evolution occurs in the states of the monocation, the double-ionization process must be sequential. The presence of sequential double ionization is supported by an angular correlation analysis of the two electrons produced in coincidence with the dissociating ions, which shows almost no correlation between the electron momenta, in contradiction to results from non-sequential ionization \cite{cheng_electron_2019,becker_theories_2012}. 
Specifically, the correlation between the $x$-component of momentum in the lab-frame of each of the two electrons produced in coincidence with D$^+$/OD$^+$ was examined. This analysis yielded a correlation coefficient of $r$ = 0.05 with a statistical uncertainty of 0.01.

Spectral analysis of the electrons emitted from an isotropic ensemble of neutral D$_2$O has shown that the first tunnel ionization preferentially populates the doublet ground state of the monocation (D$_0$) by removing an electron from the b$_1$ HOMO \cite{cheng_momentum-resolved_2020}. The molecular plane is also the nodal plane for the b$_1$ orbital, so the monocations formed in this way are expected to be preferentially aligned with the molecular plane perpendicular to the polarization \cite{muth-bohm_suppressed_2000} as was confirmed in computations presented in Ref.~\cite{cheng_momentum-resolved_2020}. The equilibrium geometry of the D$_0$ state is very near to that of the neutral ground state (X), so rearrangement of internuclear geometry should be minimal when populating this state \cite{leclerc_excited_1974}.

Likewise, removal of an a$_1$ electron from the HOMO-1 populates the first excited state of the monocation (D$_1$), leading to rapid unbending since this state has an equilibrium bend angle, $\theta_{\mathrm{HOH}}$, of 180$^\circ$ \cite{leclerc_excited_1974,jayachander_rao_sub-femtosecond_2015}.
The distinction between D$_{0}$ and D$_{1}$, however, is blurred by the presence of the laser. With a central wavelength of 800 nm, the laser is nearly resonant with the D$_{0}$ to D$_{1}$ transition \cite{leclerc_excited_1974,jayachander_rao_sub-femtosecond_2015}. Strong-field coupling between D$_{0}$ and D$_{1}$ in this intensity regime should lead to bond-softening distortions in the shape of the D$_{0}$ potential, and to rapid unbending \cite{bucksbaum_softening_1990,rottke_multiphoton_1998}.
The timescale is set by roughly half the period of the unbending mode of the D$_{1}$ monocation: $\sim$19~fs for H$_2$O$^+$ and HOD$^+$, and $\sim$26~fs for D$_2$O$^+$ \cite{reutt_molecular_1986}. In the short-pulse experiment, the monocation spends only a fraction of 10~fs in the field of the laser before undergoing further ionization followed by rapid dissociation. It is therefore unlikely that the molecule has sufficient time to substantially unbend before dissociation. By contrast, in the long-pulse experiment, full bending dynamics are possible.

\begin{figure}
    \centering
    \includegraphics[width=8.6cm, trim={0.25cm 0.5cm 1.5cm 0.5cm}]{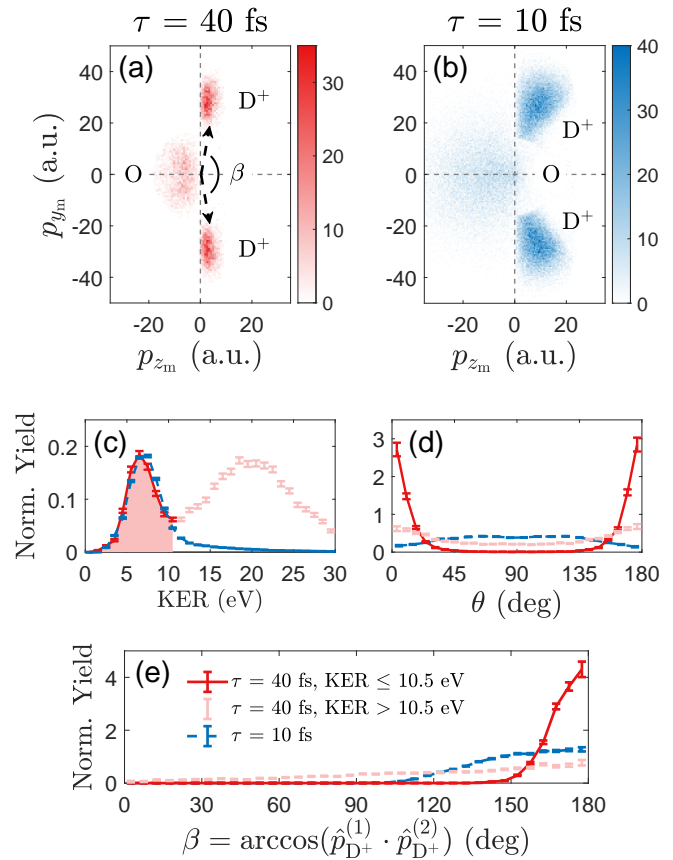}
    \caption{(a) 2D momentum distributions of the fragments in the three-body breakup channel of the heavy water dication, D$^+$/D$^+$/O, following ionization by the 40-fs pulse. Here, the bisector between the vector momenta of the two D$^+$ ions is fixed to be along the horizontal such that the vector momentum of each ion can be decomposed into a component parallel to the bisector, $p_{z_{\rm{m}}}$, and a component orthogonal to the bisector, $p_{y_{\rm{m}}}$. (b) The same as in (a) but following ionization with the 10-fs pulse. (c) The KER distribution for the case of 40~fs (solid red line) and 10~fs (dashed blue line). Light red shading indicates the KER filter used to discriminate between the D$^+$/D$^+$/O dication channel (solid red line, shaded) and contamination from the higher-KER D$^+$/D$^+$/O$^+$ trication channel (light red markers, unshaded). Here, the 40-fs data has been resized to match the peak height of the 10-fs data. (d) The distribution in angle $\theta$, as measured between the $\hat{y}_{\rm{m}}$ molecular axis and the polarization vector $\hat{\epsilon}_{\rm{laser}}$, for the case of 40~fs and 10~fs. (e) The distribution in angle $\beta$, as measured between the two D$^+$ vector momenta, for the case of 40~fs and 10~fs.}
    \label{fig:ThreeBody}
\end{figure}
 
 \subsection{\label{subsec:Three-Body}Three-Body Fragment Momentum Distributions}  
 
 The relative lack of unbending motion in the 10-fs data, as compared to the 40-fs data, is best characterized by the three-body fragmentation channel of the D$_2$O$^{2+}$ dication. In these measurements, the D$^+$/D$^+$/O channel is measured via the detection of two D$^+$ ions in coincidence, and the momentum of the undetected neutral oxygen is calculated based on momentum conservation. Defining the molecular frame in momentum-space such that the bisector of the two deuterium ions in the plane of the molecule is labeled as $p_{z_{\rm{m}}}$ \cite{streeter_dissociation_2018,reedy_dissociation_2018}, our measurements can be viewed as Figs.~\ref{fig:ThreeBody}(a) and (b) for pulse durations of 40~fs and 10~fs, respectively. Because the third fragment is left undetected, contamination from higher-charge three-body dissociations (D$^+$/D$^+$/O$^{n+}$ where $n$ is any positive integer) is possible. As seen in Fig.~\ref{fig:ThreeBody}(c), there are two distinct KER signatures in the 40-fs data, whereas in the 10-fs data, there is only one. The higher KER feature in the 40-fs data is indicative of contamination from the D$^+$/D$^+$/O$^+$ (trication) dissociation channel where a singly-charged oxygen was produced but left undetected. This was verified by directly analyzing the KER distribution in the three-body coincidence D$^+$/D$^+$/O$^+$ channel. Because it is distinct in KER, this channel can be effectively filtered out via KER, as seen in Fig.~\ref{fig:ThreeBody}(c). This KER-filtered distribution is what is utilized to produce Fig.~\ref{fig:ThreeBody}(a). As can be noted in comparing Figs.~\ref{fig:ThreeBody}(a) and (b), the 10-fs data is spread more diffusely over smaller momentum-frame bend-angles than is the 40-fs data. 
Fig.~\ref{fig:ThreeBody}(d) demonstrates that, much like the 40-fs two-body data, the 40-fs three-body data has a strong alignment preference for $\theta$~=~cos$^{-1}$($\hat{y}_{\rm{m}}\cdot\hat{\epsilon}_{\rm{laser}}$)~=~0$^\circ$. The alignment preference for the 10-fs three-body data is closer to $\theta$~=~90$^\circ$ and the distribution is substantially less sharp. Fig.~\ref{fig:ThreeBody}(e) displays the distribution over momentum-frame bend-angle, $\beta$, reaffirming the disparity in bend-angle found in comparing Figs.~\ref{fig:ThreeBody}(a) and (b).
 
 It should be noted that although $\beta$ is closely related to the geometric bend-angle of the molecule upon dissociation, $\theta_{\rm{HOH}}$, the two are distinct from each other. The equilibrium bend angle of neutral water, $\left(\theta_{\rm{HOH}}\right)_{\rm{eq.}}$, is 104.5$^{\circ}$, whereas in the short-pulse experiment the measured momentum-frame bend angle, $\beta$, ranges from $\sim$105$^{\circ}$ to 180$^{\circ}$. This disparity can be partially accounted for by some degree of unbending that occurs even in the short-pulse data. This will be explored in depth in a following publication. However, barring any internuclear motion, $\theta_{\rm{HOH}}$ and $\beta$ would be identical only if the Coulomb repulsion felt by each D$^+$ ion was exclusively along the direction of its O-D bond. Due to the distribution of charge in the water dication, this is not necessarily the case. For dissociations of higher-charge states in water (H$^+$/H$^+$/O$^{n+}$), the momentum-frame bend angle more accurately reflects the geometric bend-angle \cite{sanderson_geometry_1999,legare_laser_2005}.

 In addition to launching unbending motion, ionizing with peak intensities in the hundreds of TW/cm$^2$ can torque the molecule into alignment with the laser polarization if there exists any anisotropy in the molecule's polarizability. Whereas the neutral ground state of water is approximately isotropic in its polarizability, the D$_0$ and D$_1$ states of the monocation are not \cite{johnson_computational_2002}. Recent experimental work has demonstrated that doubly ionizing water with 40-fs pulses at 600~TW/cm$^2$ induces dynamic alignment such that the $\hat{y}_{\rm{m}}$ molecular axis preferentially aligns parallel to the laser polarization prior to dissociation \cite{mccracken_ionization_2020}. The characteristic effect of such dynamic alignment in either the three-body D$^+$/D$^+$/O or the two-body D$^+$/OD$^+$ channel is a peak in dissociation along the laser polarization axis. Looking to Fig.~\ref{fig:ThreeBody}(d), such a peak is evident when ionizing with 40-fs pulses. However, this peak is unambiguously missing for the case of a 10-fs pulse. The same is true when considering the two-body dissociation, whereby Fig.~\ref{fig:Schematic}(e) shows a peak in fragment momenta streaming along the polarization for the case of 40-fs and not for 10-fs. This suggests there is insufficient time within a 10-fs pulse for significant alignment to the polarization axis before dissociation.
 
\begin{figure}
    \centering
    \includegraphics[width=8.6cm, trim={0.125cm 0 0.125cm 0.5cm}]{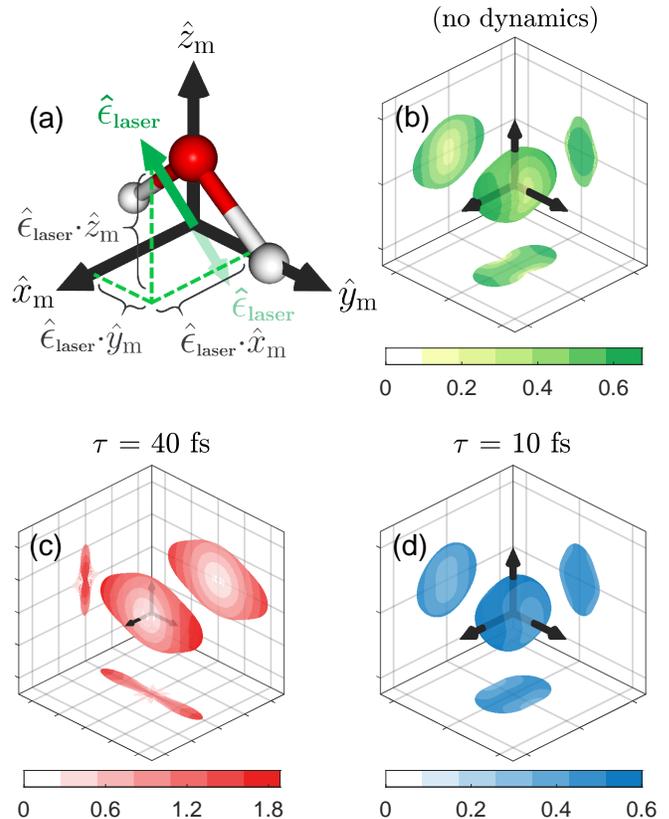}
    \caption{(a) A schematic drawing of a definite laser polarization, $\hat{\epsilon}_\mathrm{laser}$, plotted in the molecular frame with each of its projections along the molecular axes ($\hat{x}_\mathrm{m}$/$\hat{y}_\mathrm{m}$/$\hat{z}_\mathrm{m}$) labeled. (b) The angle-dependent ionization yield for D$_2$O$^+$ calculated via TD-RIS and represented as a probability distribution of the laser polarization plotted in the molecular frame as seen in (a). (c) Simulations of the alignment distribution of D$_2$O$^+$ after evolving in a strong field from $t$~=~$-\tau$/2 to $t$~=~$+\tau$/2, where $t$~=~0 is the time at which the molecule experiences peak field intensity, I$_0$~=~600 TW/cm$^2$, and $\tau$~=~40~fs is the ionizing pulse duration (FWHM). As in (b), this distribution is represented by the probability distribution of the laser polarization in the molecular frame as seen in (a). (d) The same as in (c) but for $\tau$~=~10~fs and I$_0$~=~400 TW/cm$^2$.}
    \label{fig:Alignment}
\end{figure}
 
 \subsection{\label{subsec:Simulation}Alignment Simulations}  
 
 Further evidence for the importance of dynamic alignment\textemdash particularly its dominance over geometric alignment at longer pulse durations\textemdash can be found by turning to simulation. Fig.~\ref{fig:Alignment}(b) displays the expected distribution of the laser polarization in the molecular frame directly following the first-ionization event and formation of a 79$\%$ D$_0$ and 21$\%$ D$_1$ mixture of states of the monocation. 
 This distribution was calculated via Time-Dependent Resolution in Ionic States (TD-RIS), a method of solving the time-dependent Schrodinger equation for a molecule in a strong laser field \cite{spanner_one-electron_2009,spanner_dyson_2012,spanner_molecular_2013}. 
 The method, as applied to water, has been discussed in detail in a previous work \cite{cheng_momentum-resolved_2020}. The TD-RIS distribution is computed by solving the electronic ionization dynamics for fixed nuclei and hence includes no nuclear dynamics following the ionization event; it is purely reflective of geometric alignment effects in the water monocation. Here, the polarization is predominantly distributed along the $\hat{x}_\mathrm{m}$ molecular axis. This result therefore agrees with the intuitive picture represented schematically in Fig.~\ref{fig:Schematic}(a).
 
 To explore dynamic alignment effects, these TD-RIS distributions were augmented by rigid-rotor rotational wave packet simulations on the D$_0$/D$_1$ mixture of monocations in the presence of a strong field, assuming the geometry of the neutral ground state \cite{makhija_metric_2012}. 
 The goal is to estimate the dynamic alignment of the monocation immediately prior to dication formation. Since the polarizability anisotropy of D$_0$ and D$_1$ are nearly identical, we simulate rotational dynamics using the polarizability tensor of the D$_0$ state and inertia tensor of the neutral ground state of D$_2$O~\cite{johnson_iii_nist_2020}. A rotational wavepacket is initialized with the molecular axis distribution $P(\theta,\chi)$ consistent with the TD-RIS ionization probability distribution in Fig.~\ref{fig:Alignment}(a). Here, the Euler angles $\theta$ and $\chi$ track the alignment of the D-D axis to the laser polarization in the lab frame, and the rotation of the oxygen atom around this axis, respectively. 
 To achieve this, the method in Ref.~\cite{lytova_n_2_2020} applied to N$_2$ is adapted to the asymmetric top D$_2$O. 
 The matrix representation of the amplitude $A(\theta,\chi)=\sqrt{P(\theta,\chi)}$ is constructed in the symmetric top basis, by expressing  $A(\theta,\chi)$ as an expansion of the Wigner matrix elements $D^K_{QS}(\phi, \theta, \chi)$. Euler angles $\theta$ and $\chi$ correspond to the spherical azimuthal and polar angles of the laser polarization in the molecular frame plots of Fig.~\ref{fig:Alignment}~\cite{makhija_orientation_2016}. All quantities are independent of the Euler angle $\phi$ since the Hamiltonian is independent of this angle. The functions $D^K_{QS}(\phi, \theta, \chi)$ span the space of orientations with $K= 0,1,2...$, and both $Q$ and $S$ ranging from $-K$ to $K$~\cite{zare_angular_1988}. Symmetries of the interaction Hamiltonian restrict the expansion of $A(\theta,\chi)$ to terms with even $K$ and $S$, and require that $Q=0$~\cite{makhija_orientation_2016}. The matrix elements of the $D^K_{QS}(\phi, \theta, \chi)$ in the symmetric top basis can be expressed in terms of the Wigner 3j Symbols~\cite{zare_angular_1988}, thus providing the matrix representation of $A(\theta,\chi)$ up to a finite order in $K$. We find that the $K = 2$ terms dominate the expansion of $A(\theta,\chi)$, and we include terms up to $K = 8$. 
 Each thermally populated neutral rotational state at a temperature of 30~K is transformed to the corresponding ion rotational state using this $A$ matrix, and calculation of the subsequent rotational dynamics proceed as described in prior publications~\cite{seideman_nonadiabatic_2005,rouzee_field-free_2008,pabst_computational_2010,makhija_metric_2012}.
 
 Simulating the rotational dynamics in the monocation for both long and short pulses yield drastically different results. Figs.~\ref{fig:Alignment}(c) and (d) display the final orientation of the monocations with respect to the polarization after time-evolution in a 40-fs and 10-fs pulse, respectively. Although it is the molecule's orientation with respect to the polarization that evolves in time (and not the orientation of the polarization itself), these distributions are plotted in the same way as Fig.~\ref{fig:Alignment}(b), displaying the expected distribution of the polarization in the molecular frame after time-evolution. In comparing Figs.~\ref{fig:Alignment}(c) and (d), we note that the orientation of the polarization vector following alignment within a 10-fs pulse is largely unchanged from the initial distribution displayed in Fig.~\ref{fig:Alignment}(b). Here the polarization is still primarily aligned along $\hat{x}_\mathrm{m}$. By contrast, the orientation of the polarization following alignment within a 40-fs pulse is markedly different. This distribution is predominantly aligned along $\hat{y}_\mathrm{m}$.
 
 \subsection{\label{subsec:Separating} Identifying the Contributions of Geometric and Dynamic Alignment}  
 
 In the short pulse experiment, we now demonstrate that the angular distribution of fragment ions is largely a consequence of geometric alignment. As a result of our TD-RIS calculations, we estimate that the majority ($\sim$79$\%$) of monocations are formed in the D$_0$~(1b$_1$)$^{-1}$ state. Shown schematically in Fig.~\ref{fig:Schematic}(a) and numerically in Fig.~\ref{fig:Alignment}(b), geometric alignment dictates that the removal of an electron from the HOMO will predominantly occur with the polarization parallel to the $\hat{x}_\mathrm{m}$ molecular axis. A more detailed analysis of the angle-dependent ion yields supports this interpretation \cite{cheng_momentum-resolved_2020}. In forming the dication, if the molecule undergoes minimal changes to its internuclear geometry and alignment during the pulse, we should expect the second ionization to occur at the same geometry and alignment as the first. This pathway, X~$\to$~D$_0$~(1b$_1$)$^{-1}\to$~S$_0$~(1b$_1$)$^{-2}$, may therefore be the dominant pathway when producing D$^+$/OD$^+$ fragmentation following the sequential double ionization of water using 10-fs pulses. 
 
 Further evidence of specifically populating S$_0$ can be found by looking to electron spectral analysis as well as theoretical modelling of the ions' KER. Electrons captured in coincidence with this two-body decay channel of the water dication have an energy that is characteristic of predominantly ionizing from the HOMO \cite{cheng_momentum-resolved_2020}. Separately, according to dissociation simulations of the water dication from the various electronic states \cite{gervais_h2o2_2009}, the mean value of the KER distribution when dissociating from the S$_0$ state is approximately 6.15~eV, which agrees very well with the experimental value measured here for the case of $\tau$~=~10~fs: 6.17~eV (see Fig.~\ref{fig:Schematic}(c)).
  
  As for the dissociating ions' angular dependence: we will henceforth consider the two-body dissociation of D$_2$O$^{2+}$ to occur roughly along the D-D axis (along $\hat{y}_\mathrm{m}$). While this is not exactly accurate outside of the limit where $\beta~\to$~180$^\circ$, this provides an approximate and convenient way to quantify the ions' alignment with respect to the polarization. Looking to our 10-fs two-body data, as seen in Fig.~\ref{fig:Schematic}(c), this interpretation implies that the polarization is predominantly orthogonal to $\hat{y}_\mathrm{m}$. Therefore, a pathway in which the molecule remains with $\hat{\epsilon}_\mathrm{laser}$ parallel to $\hat{x}_\mathrm{m}$, such as X~$\to$~D$_0$~(1b$_1$)$^{-1}$~$\to$~S$_0$~(1b$_1$)$^{-2}$, is well supported by our experimental observations. However, as seen in Fig.~$\ref{fig:Schematic}$(a), all states involving ionization from the HOMO~(1b$_1$) or \mbox{HOMO-1}~(3a$_1$) are preferentially formed when either $\hat{\epsilon}_\mathrm{laser}$ is along the $\hat{x}_{\mathrm{m}}$ or $\hat{z}_{\mathrm{m}}$ direction. Therefore, even those pathways that are preferentially formed at some intermediate orientation between $\hat{x}_{\mathrm{m}}$ and $\hat{z}_{\mathrm{m}}$, such as X~$\to$~D$_0$~(1b$_1$)$^{-1}$~$\to$~T$_0$~(1b$_1$)$^{-1}$(3a$_1$)$^{-1}$ would appear in the ion angular distribution as perpendicular to the polarization. These pathways may also contribute to this two-body breakup channel.
 
  We now turn to the long-pulse experiment, wherein the effects of dynamic alignment and internuclear motion dominate over geometric alignment. With a pulse duration of 40~fs and peak intensity of 600~TW/cm$^2$, we expect a significant amount of unbending and re-alignment to occur during the pulse. As seen in Fig.~\ref{fig:ThreeBody}(a) and (e), the three-body breakup of D$_2$O$^{2+}$ following long-pulse ionization demonstrates substantial unbending motion, with $\beta$ ranging from $\sim$145$^\circ$ to 180$^\circ$. Additionally, both the two-body data (see Fig.~\ref{fig:Schematic}(e)) and three-body data (see Fig.~\ref{fig:ThreeBody}(d)) indicate a large degree of alignment of the $\hat{y}_\mathrm{m}$ to the laser polarization, with $\theta$ sharply peaked around 0 and 180$^\circ$.
  
  As indicated by Fig. \ref{fig:Alignment}(c), allowing the D$_2$O$^+$ monocation to evolve within a $\tau$~=~40~fs pulse from $t$~=~$-\tau/2$ to $t$~=~$+\tau/2$ results in a significant degree of final alignment along $\hat{y}_\mathrm{m}$. However, as we expect the majority of second ionization events to occur around the peak of the pulse ($t$~=~0), the simulated timescale may seem unrealistically long. Here we offer a plausible explanation for more rapid alignment by considering unbending and alignment as coupled actions. In a purely classical picture, one can imagine a small subset of D$_2$O$^+$ ions in which one of the two O-D bonds is already initially oriented along the polarization direction. A favorable alignment can be reached not only by rotating the molecule around the center of mass, but by rotating a single deuterium about the oxygen by $\sim$75$^\circ$, simultaneously unbending and aligning the molecule. With rotation and unbending occurring simultaneously, alignment would occur on a faster timescale than what is suggested by Fig.~\ref{fig:Alignment}(c), as this simulation neglects any and all unbending motion. Additionally, recent simulations of the sequential ionization of water with 20-fs pulses have demonstrated that continued unbending motion in the dication is expected prior to dissociation \cite{koh_ionization_2020}. Therefore coupled unbending and alignment is expected to continue even after the formation of the dication prior to dissociation.
  
  
\begin{figure}
    \centering
    \includegraphics[width=8.6cm, trim={0cm 0.5cm 0.5cm 0}]{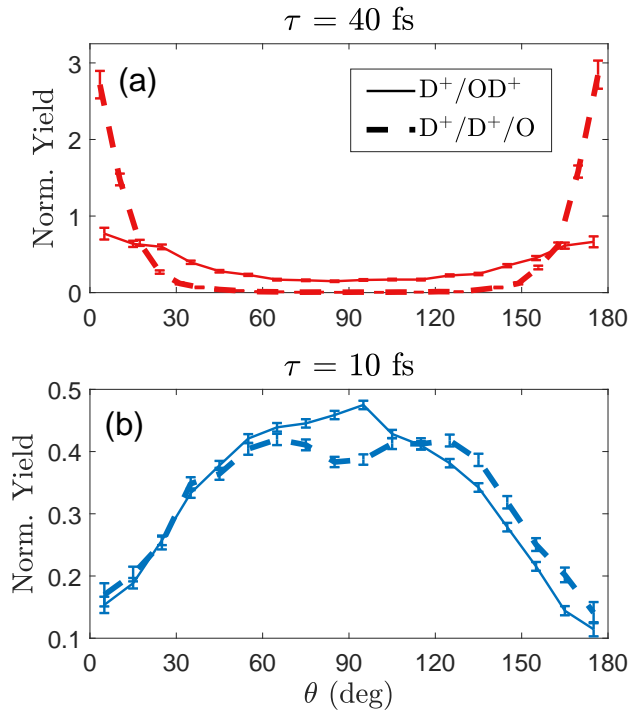}
    \caption{The angular distribution of both two-body and three-body decay channels, as measured by the angle between the assumed $\hat{y}_{\mathrm{m}}$ molecular axis and the laser polarization, $\epsilon_{\rm{laser}}$. For both (a) the long-pulse data (in red)  and (b) the short-pulse data (in blue), the two-body dissociation channel (D$^+$/OD$^+$) is plotted as a thin solid line and the three-body dissociation channel (D$^+$/D$^+$/O) is plotted as a thick dashed line. For the two-body channel, $\theta$ represents the angle between the dissociation axis and the laser polarization. For the three-body channel, $\theta$ = arccos($\hat{\epsilon}_{\rm{laser}} \cdot \hat{y}_m$).}
    \label{fig:2to3Body}
\end{figure}
  
  The two-body and three-body  decay channels correlate with different states of the dication: the lower-lying states, such as the T$_0$ (1b$_1$)$^{-1}$(3a$_1$)$^{-1}$ and S$_0$ (1b$_1$)$^{-2}$, predominantly decay into two bodies, whereas three-body decay occurs predominantly for higher-lying states of the dication, such as the T$_1$ (1b$_2$)$^{-1}$(1b$_1$)$^{-1}$, T$_2$ (1b$_2$)$^{-1}$(3a$_1$)$^{-1}$, and S$_3$ (1b$_2$)$^{-1}$(1b$_1$)$^{-1}$ \cite{gervais_h2o2_2009}. Each of these higher-lying states has a vacancy in the HOMO-2 (1b$_2$) orbital. The rules of geometric alignment imply that ionization to these higher-lying states (and subsequent dissociation into three bodies) is enhanced when $\hat{\epsilon}_\mathrm{laser}$ is along $\hat{y}_\mathrm{m}$,
  so direct comparisons between the two-body and three-body angular distributions should reflect this enhancement. 
  An enhancement along $\theta$~=~0~and~180$^\circ$ can be seen in Fig. \ref{fig:2to3Body}(a) in the three-body channel of the long-pulse experiment compared to the two-body channel. 
  For the same reason, in Fig.~\ref{fig:2to3Body}(b) a more subtle depletion of ion yield at $\theta$~=~90$^\circ$ can been seen in the short-pulse three-body data compared to the two-body data. 
  Thus the long-pulse data, though dominated by dynamic alignment, still bear signatures from geometric alignment.
  
  The simple scenario suggested by Fig.~\ref{fig:Alignment}(c), wherein long-pulse dynamics in the monocation drive the $\hat{y}_\mathrm{m}$ of D$_2$O$^+$ into alignment with the polarization before ionization to the dication state, still needs some modification, because it is not consistent with the predictions of geometric alignment. In this picture, the molecule is aligned with $\hat{y}_\mathrm{m}$ parallel to $\hat{\epsilon}_{\mathrm{laser}}$ upon formation of the dication. According to geometric alignment, this orientation should correspond to a minimum in ionization yield, since the lower-lying states all prefer alignment along either $\hat{x}_\mathrm{m}$ or $\hat{z}_\mathrm{m}$. However we observe a clear maximum in ionization yield at this orientation. This inconsistency may be largely alleviated if we assume that unbending and alignment continue to occur in the dication as well as the monocation. In this picture, the dication may be preferentially formed when the molecule is still bent and unaligned (with either $\hat{x}_\mathrm{m}$ or $\hat{z}_\mathrm{m}$ along the polarization), but subsequently unbends and aligns such that $\hat{y}_\mathrm{m}$ is largely along $\hat{\epsilon}_{\mathrm{laser}}$ during dissociation \cite{koh_ionization_2020,tong_post_2005}.
  
  Further evidence for unbending and alignment dynamics in the dication can be found in Fig.~\ref{fig:ThreeBody}(c), which shows that the majority of three-body D$^+$/D$^+$ coincidences are not from the D$^+$/D$^+$/O dication channel but rather from the D$^+$/D$^+$/O$^+$ trication channel with a significantly higher KER signature. This trication contamination is not present in the 10-fs data, suggesting the presence of enhanced multiple ionization in the 40-fs data \cite{mccracken_geometric_2017,liu_charge_2015}. The dication D$^+$/D$^+$/O channel may simply be the remnant dications that were not ionized further during the 40-fs laser pulse. This strong-field enhanced multiple ionization has been observed to be most prevalent for water aligned with the $\hat{y}_\mathrm{m}$ molecular axis along the polarization \cite{koh_ionization_2020,mccracken_geometric_2017,mccracken_ionization_2020}. 
  
  In fact, the majority of unbending and alignment may occur in the dication and not the monocation. Evidence for this can be seen in Figs.~\ref{fig:ThreeBody}(d) and (e), comparing the dissociation geometries of the dication (KER $\leq$ 10.5 eV) and trication (KER $>$ 10.5 eV) three-body decays in the 40-fs data. Here the dication is more sharply aligned with the field and $\beta$ is more sharply peaked at 180$^\circ$, indicating a more linear geometry. As the molecules unbend, many may be ionized to a trication state whereupon they undergo immediate Coulomb explosion while still bent and only partially aligned. Any dications that remain continue to align and unbend before finally dissociating in the three-body channel shown in Fig. \ref{fig:ThreeBody}(a). As seen in Fig.~\ref{fig:ThreeBody}(e), the trication signal is spread over a broad distribution of bend-angles, which indicates rapid motion along the bend-angle coordinate in the dication during the formation of the trication.

\begin{figure}
    \centering
    \includegraphics[width=8.6cm, trim={0.125cm 0.5cm 1.0cm 0.1cm}]{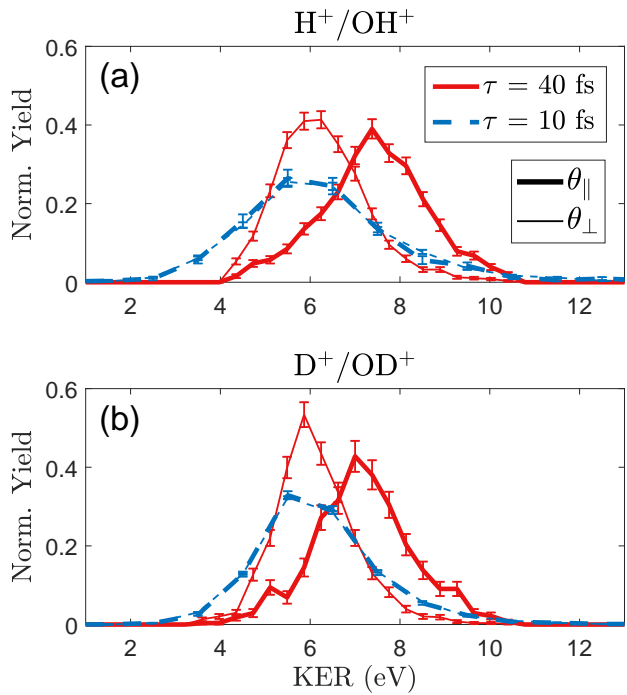}
    \caption{The KER distribution for the two-body decay pathway of doubly-ionized water H$_2$O$^{2+}$~$\to$~H$^+$/OH$^+$ (a), and deuterated water, D$_2$O$^{2+}$~$\to$~D$^+$/OD$^+$ (b). In each case, the 40-fs data is plotted in solid red lines, and the 10-fs data is plotted in dashed blue lines. Both long and short-pulse data are further subdivided between dissociation aligned along the polarization direction: $\theta_\parallel$~=~0$^\circ$~$\pm$~30$^\circ$ (thick lines) and dissociation aligned perpendicular to the polarization direction: $\theta_\bot$~=~90$^\circ$~$\pm$~30$^\circ$ (thin lines). All distributions are independently normalized to integrate to one.}
    \label{fig:AngleKERs}
\end{figure}
  
  Our final observation in the comparison of these two data sets is a clear disparity in KER. As seen in Fig.~\ref{fig:Schematic}(d), the long-pulse two-body KER is noticeably upshifted when compared to the short-pulse two-body KER. This effect is further elucidated by separating out the long-pulse two-body data into two distinct bins: dissociations that occurred parallel to the polarization, in which $\theta_\parallel$~=~0$^\circ$~$\pm$~30$^\circ$ (the majority of counts), and those that occurred perpendicular, in which $\theta_\bot$~=~90$^\circ$ $\pm$~30$^\circ$. This yields the curves shown in Figs.~\ref{fig:AngleKERs}(a) and (b) for H$_2$O and D$_2$O, respectively. As seen in either panel, the short-pulse KER remains largely unchanged between these two binnings (the two curves lie on top of each other), whereas in the long-pulse experiment the KER measured for parallel dissociations is centered significantly ($\sim$1.5~eV) higher than the KER for perpendicular dissociations.
  This disparity in KER between parallel and perpendicular dissociations is not merely due to the preferential formation of different electronic states at different orientations with respect to the polarization. If it were, there would exist the same disparity in the short-pulse data. This suggests that coupled rotations, bending, and stretching dynamics, or depletion due to further ionization, all could give rise to this shift in KER, but the precise mechanism is not presently known. Models that predict the KER observed in this experiment may therefore need to include sequential ionization and dissociation with simultaneous unbending and alignment of both the monocation and dication. We know of no work that has attempted such. What we can say definitively is that this phenomenon is reproducible across the varying isotopes of water. 
  As seen by comparing Fig. \ref{fig:AngleKERs}(a) and (b), the largest discrepancy between the two isotopes is, predictably, that the D$_2$O wavepacket---on account of its larger mass---is more sharply peaked and more localized in energy-space than is the H$_2$O wavepacket. This observation holds in comparing each of the 4 curves between the two panels.

\section{\label{sec:level5}Conclusions}

Close analysis of the two- and three-body fragmentation channels that follow the double ionization of water demonstrate that the physical mechanisms of SFI and fragmentation have a strong dependence on the ionizing pulse duration in the 10-fs to 40-fs range. This has a profound effect on coulomb-explosion imaging.  We have further demonstrated how certain nuclear dynamics in the sequential double-ionization of water can be minimized by ionizing with a sufficiently short pulse duration. In particular, when doubly ionizing water with 40-fs pulses, there appear to be strong distortions to the molecule's internuclear geometry and alignment with respect to the polarization. This is consistent with the distortions observed in previous work on the water dication \cite{mccracken_geometric_2017,mccracken_ionization_2020} as well as the higher-charge states of water \cite{sanderson_geometry_1999,legare_laser_2005}. These effects manifest themselves before dissociation and thus complicate the interpretation of angle-resolved ion yields. However, when ionizing with 10-fs pulses, changes to the molecule's internuclear geometry and alignment are drastically reduced, and angle-resolved ion yields can be interpreted using a greatly simplified model of tunnel ionization. Further detail on the key differences between short- and long-pulse regimes could follow from more comprehensive modeling of these ultrafast processes as well as experiments with more intermediate pulse durations that span the gap between the two pulse durations considered here. 

\begin{acknowledgments}
AJH, RF, GAM and PHB were supported by the National Science Foundation. AJH was additionally supported under a Stanford Graduate Fellowship as the 2019 Albion Walter Hewlett Fellow. CC and TW gratefully acknowledge support from the Department of Energy under Award No. DE-FG02-08ER15984. VM and HM were supported by the Summer Science Institute at University of Mary Washington.
\end{acknowledgments}

\bibliography{main}

\end{document}